\documentclass[twoside,showpacs,superscriptaddress,twocolumn,floatfix,a4paper,aps,pra]{revtex4}

\usepackage{color}
\usepackage{graphicx}
\usepackage[utf8x]{inputenc}
\usepackage[T1]{fontenc}
\usepackage{siunitx}
\usepackage{amstext}
\usepackage{amsmath}
\usepackage{hyperref}
\usepackage{dcolumn}

\newcommand\ket[1]{|#1\rangle}
\newcommand\bra[1]{\langle #1|}

\begin{document}

\title{Experimental measurement of the collectibility of two-qubit states}

\author{Karel Lemr}
\email{k.lemr@upol.cz}
\affiliation{RCPTM, Joint Laboratory of Optics of Palacký University and Institute of Physics of Czech Academy of Sciences, 17. listopadu 12, 771 46 Olomouc, Czech Republic}

\author{Karol Bartkiewicz} \email{bartkiewicz@jointlab.upol.cz}
\affiliation{Faculty of Physics, Adam Mickiewicz University,
PL-61-614 Pozna\'n, Poland}
\affiliation{RCPTM, Joint Laboratory of Optics of Palacký University and Institute of Physics of Czech Academy of Sciences, 17. listopadu 12, 771 46 Olomouc, Czech Republic}

\author{Antonín Černoch} \email{acernoch@fzu.cz}
\affiliation{Institute of Physics of the Czech Academy of Sciences, Joint Laboratory of Optics of PU and IP AS CR, 
   17. listopadu 50A, 772 07 Olomouc, Czech Republic}

\date{\today}

\begin{abstract}
We present a proof-of-principle experiment demonstrating measurement of 
the collectibility, a nonlinear entanglement witness proposed by Rudnicki {\it et al.} 
[Phys. Rev. Lett. {\bf 107}, 150502 (2011)]. 
This entanglement witness works for both mixed
and pure two-qubit states. In the later case
it can be used to measure entanglement in terms of the negativity.
We measured the collectibility for three distinct classes 
of photonic polarization-encoded two-qubit states, i.e.,
maximally entangled,  separable and maximally mixed states. 
We demonstrate that the measurement procedure is
feasible and robust against typical experimental shortcomings 
such as imperfect two-photon indistinguishability.
\end{abstract}


\maketitle

\section{Introduction}
Quantum entanglement is a particularly 
intriguing phenomenon~\cite{Mintert,H4a}. 
Since its conception in the famous EPR paper~\cite{EPR}, 
it has received a great deal of interest. 
Over the years both theoretical and experimental research 
was dedicated to its investigation~\cite{Plenio,H4a}.

Two distinct strategies are usually applied to quantum 
entanglement detection and quantification. 
The first approach to entanglement characterization is based 
on quantum state tomography and density matrix estimation~\cite{Salles}. 
Knowing the state's density matrix, it is possible to apply 
various entanglement criteria and entanglement measures 
(e.g., the Peres-Horodecki citerion~\cite{Simon}). 
Performing full state tomography is however 
a lengthy procedure since the number of required measurements 
grows exponentially with the dimension of Hilbert space.

The second strategy is based on so-called entanglement 
witnesses (EW). Measuring a simple linear EW
involves performing suitable local measurements. 
Their correlation across involved parties then reveals 
the entanglement [e.g. Clauser-Horne-Shimony-Holt (CHSH) experiment \cite{CHSH}]. 
This strategy, however, requires some a priory information 
about the investigated state to properly choose 
the performed measurements. 
A revised  version of this approach allows to 
estimate the amount of the available entanglement
from the maximal amount of the CHSH inequality 
violation~\cite{EntFromCHSH}. 

One can describe entanglement 
using collective (nonlinear) entanglement witnesses~\cite{Horodecki03}
that depend on joint measurements on several copies 
of the entangled state.
As demonstrated experimentally, e.g., by Bovio {\it et al.} ~\cite{Bovino}, 
it is possible to apply a specific two-copy witness
to detect quantum entanglement.
In this way one can also measure entanglement in terms 
of concurrence~\cite{MintertPRL,ZHChen}
(related to the entanglement of formation)
or other two-copy witnesses~\cite{Aolita,Walborn}.
There is a number of other nonlinear entanglement witnesses 
and quantifiers that do not rely on quantum state tomography
~\cite{Badziag,Huber10,GRW07,EBA07,Augusiak08,OH10,
Rudnicki,JMG,Rudnicki2,RPHZ14,Sheng15}. 
In particular, adaptive approach to 
two-qubit entanglement detection was proposed
in Ref.~\cite{Park,Laskowski}.


\begin{figure}
\includegraphics[scale=0.8]{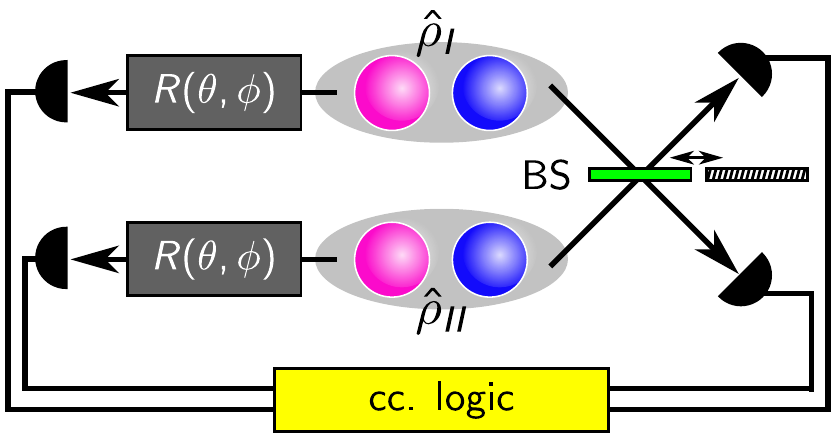}
\caption{\label{fig:concept} 
Conceptual scheme for measuring the collectibility
of polarization encoded two-qubit states
with linear optics.
Two copies of a two-qubit state 
$\hat{\rho}_I$ and $\hat{\rho}_{II}$ are prepared. 
Polarization of one the photons from each pair 
is measured locally.
At the same time, two different measurements 
on the other two photons are performed interchangeably: 
(i) simple unconditional detection (balanced beam splitter BS is removed) 
and (ii) two-photon bunching measurement corresponding 
to singlet state projection (BS inserted). 
The resulting four-fold coincidences are registered 
and the collectibility is then calculated from these 
coincidences using the procedure described in the text.}
\end{figure}

All the above mentioned approaches
have their merits, but here we will focus
on the proposal of  Rudnicki {\it et al.}~\cite{Rudnicki,Rudnicki2}. 
It makes use of suitable measurements performed simultaneously 
on two copies of the investigated quantum state (see the conceptual 
scheme in Fig.~\ref{fig:concept}).
This nonlinear entanglement witness is referred to as
the {\it collectibility}, because values of the projections of the analyzed state onto the 
set of separable states are accumulated collectively.
In a subsequent work, the authors of Ref.~\cite{Rudnicki} have 
extended the notion of the collectibility to apply also to mixed states~\cite{Rudnicki2}
and intorduced a collective nonlinear entanglement witness $W(\hat\rho)$.

This technique does not share the disadvantages of the previously mentioned methods,
{\it i.e.}, it allows to detect two-qubit entanglement for an arbitrary state
and for pure states it allows to measure the amount of entanglement
in terms of the negativity~\cite{Rudnicki2}, an entanglement 
measure corresponding to the entanglement cost under the positive partial transpose operations~\cite{NegPPT1,NegPPT2}. 
Moreover, the notion of the collectibility is based on the same 
principles as the entropic uncertainty relations, thus,
its value can have a physical interpretation.
However, a completely universal negativity measurement
requires in general four copies of the two-qubit 
state~\cite{Augusiak08,Bartkiewicz15a,Bartkiewicz15b}.
In comparison to other previously proposed methods, the collectibility requires only one two-photon interaction to be implemented.

To measure the colletibility of two-qubit systems, it is required to perform collective 
measurements on two copies of the state. 
On the platform of linear optics with qubits encoded as polarization states of individual photons, 
these measurements include various local polarization projections 
combined with two-photon interference (Hong-Ou-Mandel effect \cite{HOM}).
The set of performed measurements is state-independent 
and, thus, does not require any a priory knowledge. The use of local projections is an additional experimental benefit, because these can be simultaneously treated as heralds when using imperfect single-photon sources (e.g. those based on spontaneous frequency down-conversion).


Here, we demonstrate that the entire collectibility measurement 
procedure is experimentally feasible with current level of quantum 
optics technology. We test the Rudnicki {\it et al.} \cite{Rudnicki} 
method on three distinct two-photon quantum states (one pure entangled,
one pure separable and one maximally mixed). 
Consequently, we compare the observed values of $W(\hat{\rho})$ and their theoretical values.
We demonstrate that all the measurements can be corrected for 
typical (and unavoidable) experimental shortcomings such as 
imperfect two-photon indistinguishability. 
The possibility to calibrate the measurement apparatus is 
an important prerequisite allowing this procedure to be used 
in realistic experimental conditions.

\section{Theoretical framework}
We focus on two-qubit states $\hat{\rho}=\hat{\rho}_{A,B}$. Let us assume that qubit $A$
 is projected onto $\ket{a_+}=\cos\frac{\theta}{2}\ket{H} + e^{i\phi}\sin\frac{\theta}{2}\ket{V}$
or  $\ket{a_-}=\cos\frac{\theta}{2}\ket{V}
- e^{-i\phi}\sin\frac{\theta}{2}\ket{H}$, where $H$ and $V$ stand for a horizontally-polarized and vertically-polarized photon, respectively. 
These projections are set by the $R(\theta,\phi)$ rotations marked in Fig.~\ref{fig:concept}. The probability associated with these projections reads $p_\pm = \mathrm{tr\,}(\hat\chi_\pm)$ and the reduced state is described by the following single-qubit density matrix $\hat\sigma_\pm=\hat\chi_\pm/p_\pm$,
where $\hat\chi_\pm=\mathrm{tr\,}_A[\rho_{A,B}(\ket{a_\pm}\bra{a_{\pm}}\otimes\hat\openone)_{A,B}]$.
The collectibility for two-qubit states \cite{Rudnicki,Rudnicki2} can be calculated as 
$Y(\hat{\rho},\theta,\phi) = \frac{1}{4}\left(\sqrt{G_{+}G_{-}}+\sqrt{G_{+}G_{-}-G^2_{}} \right)^2,$
where $G_{\pm}=p_\pm\sqrt{\mathrm{tr\,}(\hat\sigma^2_\pm)}$, 
$G=\sqrt{p_+p_-\mathrm{tr\,}(\hat\sigma_+\hat\sigma_-)}$ are Gramm matrix elements  \cite{Rudnicki} that depend on probabilities $p_\pm$ and two-photon overlaps $\mathrm{tr\,}(\hat\sigma_{i} \hat\sigma_j)$ ($i,j=\pm$). These overlaps are measured  by using the following identity $\mathrm{tr\,}(\hat\sigma_i\hat\sigma_j)=\mathrm{tr\,}[\hat S(\hat\sigma_i\otimes\hat\sigma_j)],$
where $\hat S=\hat\openone\otimes\hat\openone-2\hat P^-$ and $\hat P^-$ is a projector on  polarization-encoded
singlet state. This projection  is implemented by overlapping photons $\hat\sigma_i$ and $\hat\sigma_j$
on a balanced beamsplitter and measuring
the rate of anticoalescence of the photons transmitted  through the beamsplitter.
This technique is very useful in measuring quantum properties of microsopic
systems (see, e.g., Ref.~\cite{EntFromCHSH,HOM,quantcorr,sfid,tomo1,tomo2}).
Direct calculations reveal that the maximum of collectibility does not depend on the parameter $\phi$, thus, 
we  set $\phi=0$ \cite{Rudnicki}.  Measuring the maximum value of 
$Y(\hat\rho)=Y_{\max{}} (\hat{\rho},\theta,\phi)>1/16$  allows to both detect 
and quantify the entanglement of pure two-qubit states \cite{Rudnicki}. However, for mixed states it is 
more convenient to apply a closely related  collective   witness 
$W(\hat{\rho})$   \cite{Rudnicki}, i.e.,
\begin{equation}\label{eq:W}
W(\hat{\rho})=(\eta+G^2_+ + G^2_- +2G^2-1)/2,
\end{equation}
where the Gramm matrix elements are calculated for $\ket{a_\pm}=\ket{H},\ket{V}$. This witness depends on conditional purity  $\mathrm{tr\,}(\hat\sigma^{2}_{\pm})$  of a qubit  $B$, obtained after projecting its counterpart from subsystem $A$  onto $\ket{a_\pm}.$   The parameter $\eta=8p_+p_-\sqrt{z_{+}z_{-}} +2\max[{x_{+},x_{-}]}$ depends on
 $z_{_{\pm}}=2\,\mathrm{tr\,}[\hat P^{-}(\hat\sigma_\pm^{\otimes 2})]|_{\theta=0}$
and  $x_{\pm}=2\,\mathrm{tr\,}[\hat P^{-}(\hat\sigma_\pm^{\otimes2})]|_{\theta=\pi/2}$.
Measuring negative value of  $W(\hat{\rho})$  indicates the detection of entanglement. This witness is also extendible to higher-dimensional states \cite{Rudnicki}.

\section{Experimental setup}
\begin{figure}
\includegraphics[scale=0.1]{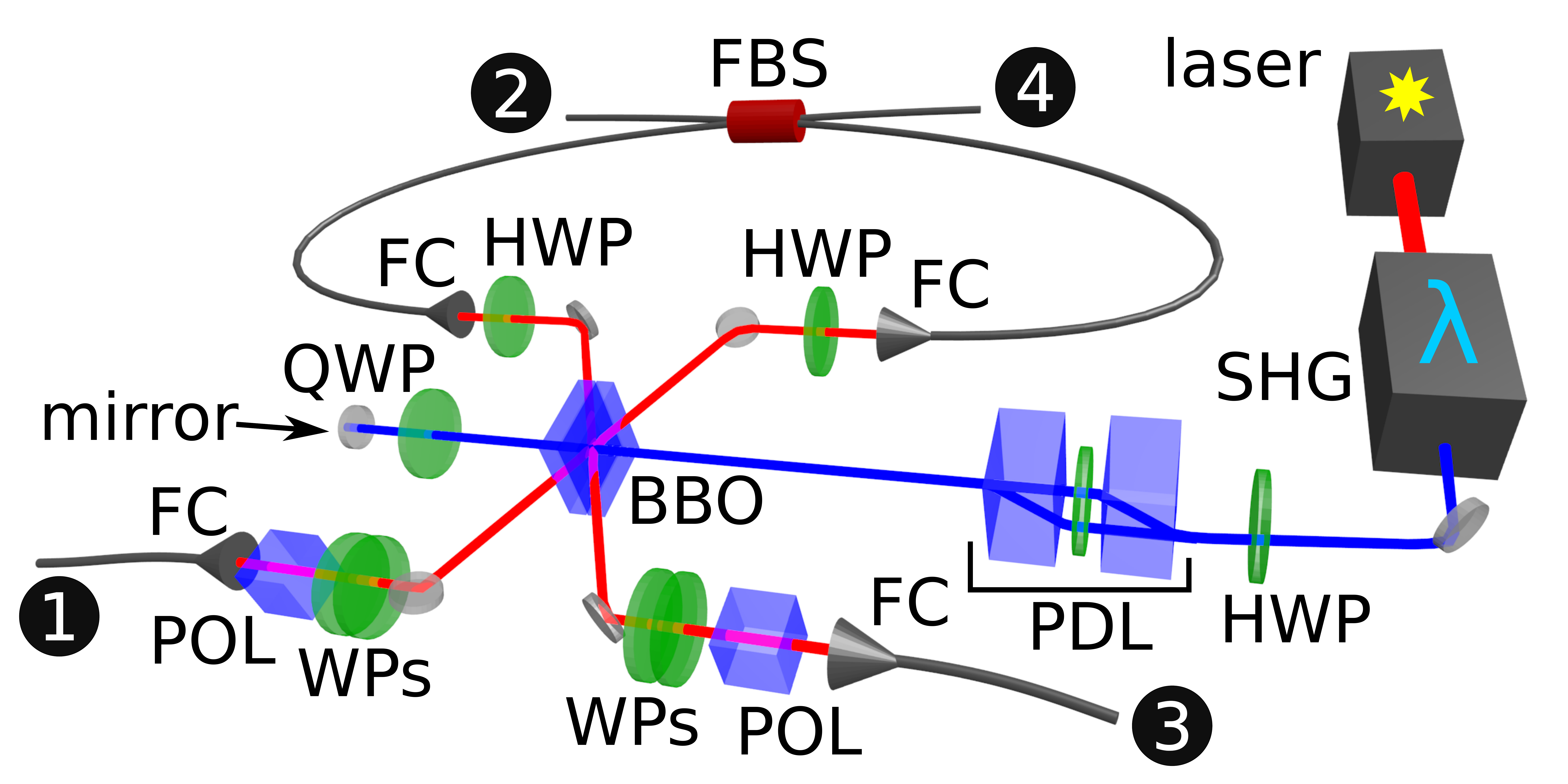}
\caption{\label{fig:setup} Experimental scheme for colectibility measurement on two copies of two-photon states. Laser source used was the Coherent Mira laser and the respective components are labeled as follows: SHG -- second harmonics generation, HWP -- half-wave plate, PDL -- polarization dispersion line (two beam displacers with one HWP in the middle), BBO -- pair of BBO crystals, QWP -- quarter-wave plate, WPs -- set of half and quarter-wave plates, POL -- polarizer, FC -- fiber coupler, FBS -- fiber beam spitter. Photon-mode numbers, as used in the text, are denoted by encircled numbers.}
\end{figure}
We have constructed a four-photon source as depicted in Fig. \ref{fig:setup}. This source is powered by femtosecond laser pulses emitted by Coherent Mira laser with \SI{76}{MHz} of repetition rate. Second harmonics generation (SHG) is then used to convert the wavelength of these pulses to \SI{413}{\nano\meter}. Typical mean power of the pumping beam at the output of SHG is about \SI{300}{mW}.

In the next step, the pumping beam is subjected to a polarization dispersion line (PDL) to correct for intrinsic polarization dispersion of the BBO crystals ($\beta$-BaB$_2$O$_4$) used to generate photon pairs. This dispersion line consists of two beam displacers and a half-wave plate placed in between. Suitable tilt of these beam displacers allows to elongate optical path of one polarization with respect to the other.

The laser beam then impinges on a BBO crystal cascade (known as the Kwiat source \cite{Kwiat}). Type I spontaneous parametric down-conversion occurring in the first and second BBO crystal coherently generates pairs of horizontally and vertically polarized photons respectively. By adjusting the pumping beam polarization, one can tune the generation rate of horizontally and vertically polarized pairs as well as their mutual phase shift. The pumping beam then propagates through a quarter-wave plate, gets reflected on a mirror, goes again through that quarter-wave plate and impinges on the BBO crystal cascade on its way back towards the source. At this point a second pair of photons is generated. The quarter-wave plate is used to rotate the pumping beam polarization so that again it compensates for the polarization dispersion in BBO crystals (now in reverse order).

Using mirrors, pairs of photons generated by both the forward and backward propagating pumping beam are directed towards fiber couplers that lead them to single-photon detectors via single-mode optical fibers. These couplers are equipped with \SI{10}{\nano\meter} interference filters in case of photons 1 and 3 and by \SI{5}{\nano\meter} interference filters in case of photons 2 and 4. Before being collected by the fiber couplers, one photon from each pair (photons 1 and 3 as depicted in Fig. \ref{fig:setup}) is subjected to polarization projection using a quarter-wave plate, a half-wave plate and a polarizing cube. The other two photons (2 and 4) are collected directly to single-mode fibers and then overlapped on a balanced fiber beam splitter (FBS) before led towards detectors. Temporal two-photon overlap on this beam splitter is adjusted by suitable choice of a fiber delay line and a motorized translation stage positioning the pumping beam back-reflecting mirror. Typical four-photon detection events occur about once per 5 minutes depending on the adjusted quantum state and polarization projection.

\section{Measurement and results}
{\it Measurement and results} -- As the first step, two copies of the investigated quantum state have to be prepared. We have performed the collectibility measurement on three different quantum states
\begin{eqnarray}\label{eq:states}
|\Psi_1\rangle & = & \frac{1}{\sqrt{2}}\left(|HV\rangle-|VH\rangle\right) \quad\mathrm{(Bell\,state),}\nonumber\\
|\Psi_2\rangle & = & |HH\rangle \quad\mathrm{(separable\,state),}\nonumber\\
{\hat\rho_3} & = & \frac{1}{4}\left(|HH\rangle\langle HH|+|HV\rangle\langle HV|+|VH\rangle\langle VH|\right.\nonumber\\
& & \left.+|VV\rangle\langle VV|\right) \quad\mathrm{(mixed\,state)}.
\end{eqnarray}

First copy of a selected state was encoded into photons 1 and 2, the second copy into photons 3 and 4. For more detailed description of the state preparation procedure see the Appendix.

\begin{table}[h]
\caption{\label{tab:results_S} Summarized results obtained for states defined in Eq.~(\ref{eq:states}). $\mathcal{P}$ denotes experimentally measured state purity, $\mathcal{P}_\mathrm{theo}$ stands for its theoretical value, $W$ is the observed value of entanglement witness and $W_\mathrm{th}$ is its theoretical value.}
\begin{ruledtabular}
\begin{tabular}{lcccc}
Quantum state   & $\mathcal{P}$ & $\mathcal{P}_\mathrm{theo}$   & $W    $       & $W_\mathrm{th}$\\\hline
Bell state              & $0.89\pm0.02$ &       1.00                                                    & $-0.21\pm0.07$  & $-0.25$\\
Separable state & $0.96\pm0.01$ &       1.00                                                    & $+0.03\pm0.04$  & $0$\\
Mixed state             & $0.27\pm0.01$ &       0.25                                                    & $+0.73\pm0.02$  & $0.75$\\
\end{tabular}
\end{ruledtabular}
\end{table}
As a preparatory measurement, purities of each investigated state were estimated by performing their polarization projections onto horizontal, vertical, diagonal and anti-diagonal polarizations. The obtained values for states defined in Eq.~(\ref{eq:states})
 are summarized in Tab.~\ref{tab:results_S}.
 The values of $W(\hat{\rho})$ were determined by measuring four-fold coincidence rates when projecting photons 1 and 3 onto combinations of horizontal ($H$), vertical ($V$) and diagonal ($D$) polarizations. At the same time photons 2 and 4 overlap on FBS. Pumping beam mirror position was adjusted accordingly to achieve this overlap. Coincidence rates obtained this way were labelled $ccA_{IJ}$ (indexes $I$ and $J$ denote any of the above mentioned polarizations on  the first and third photon respectively). Similarly the coincidence rates $ccB_{IJ}$ were measured when photons 2 and 4 were not overlapping in time on FBS (pumping beam mirror shifted out). In order to accumulate enough signal, these four-fold coincidences were aggregated for about 10 hours for each of the required settings. Due to imperfect two-photon overlap between photons 1 and 3 we have to deal with a non-removable noise in the form of parasitic coincidences denoted $ccN$ corresponding to non-interacting (non-bunching) photons even if time overlap is adjusted. We have estimated this noise level for each investigated state and recalculated the coincidence rates accordingly
\begin{eqnarray}
\overline{ccA}_{IJ} &=& ccA_{IJ} - ccN\nonumber\\
\overline{ccB}_{IJ} &=& ccB_{IJ} - ccN.
\end{eqnarray}
Their ratio $r_{IJ}$ then reads
\begin{equation}
\label{eq:ratios}
\bar{r}_{IJ} = \frac{\overline{ccA}_{IJ}}{\overline{ccB}_{IJ}} = \frac{ccA_{IJ}-ccN}{ccB_{IJ}-ccN}.
\end{equation}
This coincidence ratio corresponds to doubled singlet detection rate, because the number of coincidences $ccB_{IJ}$ corresponds  to half of the coincidences that would have been measured with FBS completely removed. More detailed description of this procedure is provided in the Appendix.

\begin{figure}
\includegraphics[scale=1]{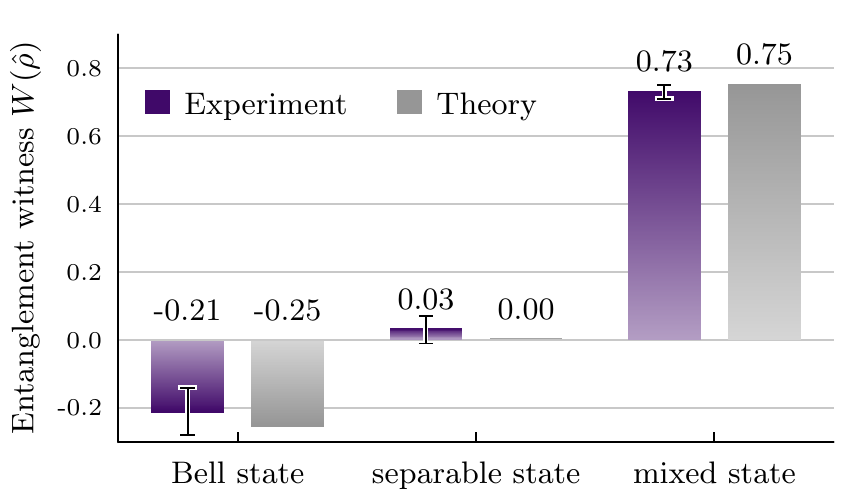}
\caption{\label{fig:W} Plotted values of the collectibility entanglement witness for the three investigated states given in Eq.~(\ref{eq:states}). Presented bars show experimentally observed and theoretically calculated values respectively.}
\end{figure}

Knowing the ratios defined in (\ref{eq:ratios}), one can express the  entanglement witness from Eq.~(\ref{eq:W})
in terms of the coincidences in the following way
\begin{eqnarray}
W(\hat{\rho}) &=& \frac{1}{2}\Bigl[\eta + \xi^2\left(1-r_{HH}\right) + \left(1-\xi\right)^2\left(1-r_{VV}\right)\nonumber\\
&  & + 2\xi\left(1-\xi\right)\left(1-r_{HV}\right) -1\Bigr],
\end{eqnarray}
where
\begin{equation}
\eta = 8\xi\left(1-\xi\right)\sqrt{r_{HH}r_{VV}}+2r_{DD}
\end{equation}
and $\xi$ is the probability of observing both photons of the same pair (either 1 and 2 or 3 and 4) horizontally polarized. For the Bell and mixed state, we have achieved to adjust $\xi = 0.50\pm0.03$ while for the separable state it was $\xi = 1.00\pm0.01$. Fig. \ref{fig:W} visualizes the theoretical and experimental values of collecitbility entanglement witness observed for states given in Eq.~(\ref{eq:states}). For exact values see Tab.~\ref{tab:results_S}.

\begin{figure}
\includegraphics[width=8cm]{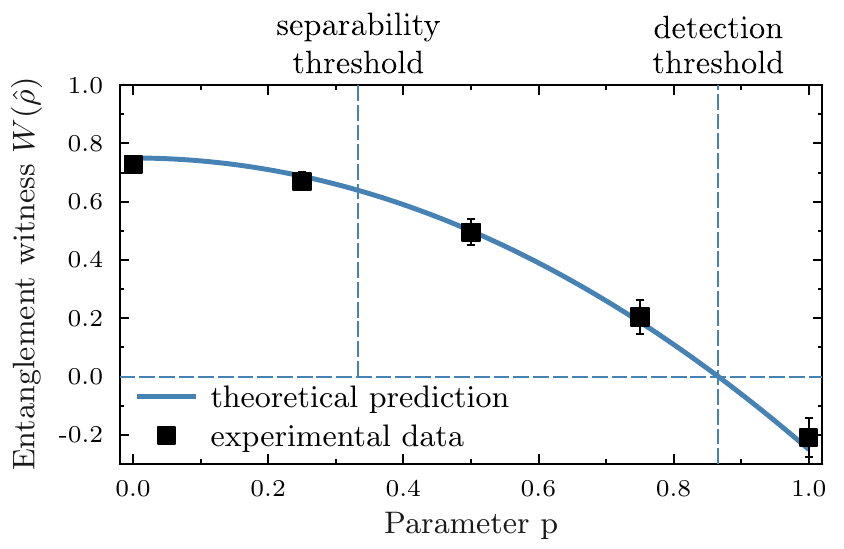}
\caption{\label{fig:Werner} The entanglement witness $W(\hat\rho)$ plotted for the Werner states defined in Eq.~(\ref{eq:Werner}). Both the separability threshold and the threshold of detectability by the collectibility witness are depicted as well.
}
\end{figure}

In the next step, we have analyzed the collectibility of Werner states which can be expressed in terms of two of the states given in Eq. (\ref{eq:states})
\begin{equation}
\label{eq:Werner}
\hat\rho_W = p |\psi_1\rangle\langle\psi_1| + (1-p) \rho_3.
\end{equation}
In order to interpolate the measurement for any value of $p\in[0,1]$, we need to average the coincidence counts (or ratios $\bar{r}_{IJ}$) for the $|\psi_1\rangle$ and $\rho_3$ state collectibility measurement with effective weights of $p^2$ and $1-p^2$ respectively. This means that with probability of $p^2$, two copies of $|\psi_1\rangle$ are prepared and their collectibility measured. Similarly, with probability of $(1-p)^2$, two copies of the $\rho_3$ are prepared. With probability of $2p(1-p)$ however, one copy of $|\psi_1\rangle$ and one copy of $\rho_3$ are inserted into the setup. Since one of these states ($\rho_3$) is a completely mixed state, the resulting measurement is identical to the case with two $\rho_3$ states. Note that both the $|\psi_1\rangle$ and $\rho_3$ states give complete random outcomes of local projections and the Hong-Ou-Mandel interference between any state and a maximally mixed state always yields identical dip depth of $1/2$. Hence we only need to mix the coincidences observed for the $|\psi_1\rangle$ and $\rho_3$ states with weights of $p^2$ and $1-p^2$ respectively. By doing so, we have been able to experimentally investigate the collectibility of Werner states as function of its parameter $p$. Note that while Werner states are entangled for any value of $p>1/3$, the collectibility is only able to detect entanglement for $p>\sqrt{3}/2\approx0.87$. Observed data as well as theoretical predictions are visualized in Fig.~\ref{fig:Werner} and summarized in Tab.~\ref{tab:resultsW_S}.
\begin{table}[h]
\caption{\label{tab:resultsW_S} Summarized results obtained for Werner states defined in Eq.~(\ref{eq:Werner}). $W$ stands for the observed value of entanglement witness and $W_\mathrm{th}$ is its theoretical prediction.}
\begin{ruledtabular}
\begin{tabular}{crr}
$p$   & $W\qquad $       & $W_\mathrm{th}$\\\hline
0.00	&$0.73\pm0.02$	&	$0.75$\\
0.25	&$0.67\pm0.03$	&	$0.69$\\
0.50	&$0.50\pm0.04$	&	$0.50$\\
0.75	&$0.20\pm0.06$	&	$0.19$\\
1.00	&$-0.21\pm0.07$	&	$-0.25$\\
\end{tabular}
\end{ruledtabular}
\end{table}

\section{Conclusions}
We have presented experimental measurement of the collective entanglement witness for three different quantum states. Our results are in quite a good agreement with the theoretically calculated values. The observed value of the collectibility witness for the Bell state ($-0.21\pm0.07$) is well below zero by about 3 multiples of its standard deviation.
%
%
In the case of the other two states, the value of $W$ is non-negative as it should be for separable states. Further to that, we were able to interpolate the collectibility for various Werner states. Experimental data support the theoretically identified relation between the collectibility witness and the Werner states parameter $p$. We therefore conclude that our experiment is a sufficient a proof-of-principle test of the collectibility as an nonlinear entanglement witness. Moreover we demonstrate how the measurement can be calibrated for experimental imperfections especially reduced two-photon overlap. Measuring collectibility requires measuring much fewer  parameters
that full quantum state tomography (QST). However, it is
time consuming due to working with SPDC-based sources of entangled photons.
Using more deterministic sources would make collectibility measurements faster and more appealing for experimentalists than QST. This especially pronounced in the case of multiqubit entangled states, where
the number of parameters needed for QST grows exponentially with the dimension of the entangled state.

\appendix*
\section{Detailed account on experimental procedures}
\subsection{Initial state preparation procedure}
We have taken the following steps in order to prepare the required two copies of investigated input state (designated $|\Psi_1\rangle, |\Psi_2\rangle$ and $\hat{\rho_3}$ in the main text). Preparation of the separable $|\Psi_2\rangle$ state is quite straightforward. In this case, half-wave plate in the pumping beam as well as the quarter-wave plate in front of the pumping beam mirror were set to zero degrees so that only one of the BBO crystals generates photons. Polarization controllers are used to maintain horizontal polarization of the photons 2 and 4 at the input of the fiber beam splitter.

A somewhat more difficult procedure has to be undertaken in order to prepare the maximally entangled state $|\Psi_1\rangle$. First, it requires inserting temporary polarizers into paths of photons 2 and 4 (before they are coupled into fibers). The pumping beam polarization was set to diagonal so that both the BBO crystals contribute to the two-photon state generation. At this point the pumping beam reflection was blocked so that only forward generated photon pairs are observed. By projecting the photon 1 and 2 onto diagonal polarization, we have adjusted the polarization dispersion line in the pumping beam. This was achieved by observing coincidence rate visibility as a function of piezo driven phase shift inside the polarization dispersion line for various setting of disbalanced paths for horizontal and vertical polarization. Once maximum contrast was obtained (about 20:1), the piezo was adjusted so to maximize the observed coincidences (this way we have set the phase in the Bell state). Afterwards, the temporary polarizer for the photon 2 was removed and the ratio between the $HH$ and $VV$ coincidences was balanced (with precision of about $5\%$) by observing coincidence rates when projecting the photon 1 onto horizontal/vertical polarization. An identical procedure was repeated to adjust the same state on the backward propagating photons 3 and 4.

For both the above mentioned pure states $|\Psi_1\rangle$ and $|\Psi_2\rangle$, the four-fold coincidences were detected by firstly observing coincident events $cc_{13}$ between photons 1 and 3 and $cc_{24}$ between photons 2 and 4. These coincidences were obtained by using time-to-amplitude modules and single channel analyzer electronics set to coincidence window of \SI{5}{\nano\second} (smaller that the laser repetition period). Subsequently, coincidence logic was used to obtain four-fold coincidences combining $cc_{13}$ and $cc_{24}$ (with coincidence window of about \SI{20}{ns}).

Maximally mixed state corresponds to the photons being completely of random polarization, mutually not correlated. By adjusting balanced generation rates from both crystals, the individual photons have this property. The correlation (and entanglement) is only visible  when photon pairs 1 and 2 or 3 and 4 are observed simultaneously. Typically, we observe about 30 000 individual photons per second and only about 500 coincident pairs. The majority of photons thus do not have their pair counterpart detected. This fact allows us to use the following procedure to obtain a mixed state. Instead of measuring the four-fold coincidence rates, we have only detected the two-fold coincidence rates $cc_{13}$ and $cc_{24}$. Since photons 1 and 3 are uncorrelated, so is their coincidence rate $cc_{13}$. The same is valid for $cc_{24}$.  Then we have just multiplied these two coincidence rates to obtain the resulting mixed state coincidence rate. Since less then 2\% of single photons result in pair coincidences, the vast majority of the signal correspond to a completely mixed state.

\subsection{Coincidence rate correction for imperfect two-photon overlap}
In an ideal case, when both the photons 1 and 3 are in the same polarization state, no coincidences $cc_{24}$ should be observed due to the two-photon bunching. In the real experiment however, there was a non-removable jitter between the generation of the first and second pair of photons due to the finite time the pulse travels through the crystals. Because of this jitter and other minor experimental imperfections, the probability of the two-photon overlap was decreased and this non-interacting portion of the photons constituted a noise. In order to process the data, we had to subtract this noise from both the $ccA_{IJ}$ and $ccB_{IJ}$. A calibration measurement was therefore performed to estimate the noise level. For all investigated states both the photons 1 and 3 were projected onto the same polarization from the set of horizontal and vertical polarizations. The obtained coincidence rates were averaged to obtain the noise level $ccN$. Corrected coincidence ratios were then calculated using the formula
\begin{equation}
{r}_{IJ} = \frac{ccA_{IJ}-ccN}{ccB_{IJ}-ccN}.
\end{equation}
Table \ref{tab:ccN} summarizes the observed noise levels for the three investigated states. Note that due to different optimal position of the fiber couplers, the noise level varies for different states. For the Bell and separable state, we have measured the noise four-fold coincidence rates while for the mixed state, we have measured the noise two-fold coincidences $cc_{24}$. That is because in the case of the mixed state, no four-fold coincidences are measured directly, but they are obtained by multiplying two-fold coincidences $cc_{13}$ with $cc_{24}$ and only the first ones need to be calibrated.

\begin{table}
\caption{\label{tab:ccN} Estimated average values of parasitic coincidences $ccN$ given for all three measured states. Presented values are relative parasitic coincidence rates with respect to overall coincidence rate observed when photons were not overlapping in time ($ccB$).}
\begin{ruledtabular}
\begin{tabular}{lcccc}
Quantum state   & relative parasitic coincidence rate $\frac{ccN}{ccB}$ \\\hline
Bell state              & $0.57\pm 0.02$\\
Separable state & $0.49\pm 0.02$\\
Mixed state             & $0.85\pm 0.01$\\
\end{tabular}
\end{ruledtabular}
\end{table}

\section*{Acknowledgements}
KL and KB acknowledge
financial support by the Czech Science Foundation under the project No. 16-10042Y and the financial support 
of the Polish National Science Centre under grant
DEC-2013/11/D/ST2/02638.
AČ acknowledges financial support by the Czech Science Foundation under the project No. P205/12/0382. The authors also acknowledge the project
No. LO1305 of the Ministry of Education, Youth and
Sports of the Czech Republic financing the infrastructure of their workplace.

\end{document}